\documentclass[aps,prd,eqsecnum,nofootinbib,superscriptaddress,preprintnumbers,11pt]{revtex4}
\usepackage{amssymb,amsmath,psfrag,epsfig,feynarts}
\usepackage{cancel}
\usepackage{subfigure}
\newcommand{\beq}{\begin{equation}}
\newcommand{\eeq}{\end{equation}}
\newcommand{\beqa}{\begin{eqnarray}}
\newcommand{\eeqa}{\end{eqnarray}}
\newcommand{\bea}{\begin{align}}
\newcommand{\eal}{\end{align}}
\newcommand{\nn}{\nonumber  \\}
\newcommand{\la}{\langle}
\newcommand{\ra}{\rangle}
\newcommand{\rar}{\rightarrow}

\def\fig#1{Fig.~{\ref{#1}}}

\begin{document}

\preprint{Brown-HET-1608} \preprint{AEI-2010-166}

\begin{center}%
{\Large\textbf{\mathversion{bold}%
Note on Bonus Relations for $\mathcal{N}=8$ Supergravity\\ Tree Amplitudes}}

\author{Song He}

\affiliation{Max-Planck-Institut f\"ur Gravitationsphysik, Albert-Einstein-Institut\\%
Am M\"uhlenberg 1, 14476 Potsdam, Germany }

\author{Dhritiman Nandan}

\affiliation{Physics Department, Brown University, Providence, Rhode
Island 02912, USA}

\author{Congkao Wen}

\affiliation{Physics Department, Brown University, Providence, Rhode
Island 02912, USA}

\maketitle

\vspace{3cm} \textbf{Abstract}\vspace{7mm}

\begin{minipage}{12.7cm}
We study the application of non-trivial relations between gravity
tree amplitudes, the bonus relations, to all tree-level amplitudes in
$\mathcal{N}=8$ supergravity. We show that the relations can be used
to simplify explicit formulae of supergravity tree amplitudes, by
reducing the known form as a sum of $(n-2)!$ permutations
obtained by solving on-shell recursion relations, to a new form as a
$(n-3)!$-permutation sum. We demonstrate the simplification by
explicit calculations of the next-to-maximally helicity
violating (NMHV) and next-to-next-to-maximally helicity violating
(N$^2$MHV) amplitudes, and provide a general pattern of bonus
coefficients for all tree-level amplitudes.

\end{minipage}

\end{center}

\newpage

\section{Introduction}

In the past several years there have been enormous progress in
unraveling the structure of scattering amplitudes in gauge theory
and gravity, such as generalized unitary-cut method at loop
level~\cite{Bern:1994zx}, and Britto-Cachazo-Feng-Witten (BCFW)
recursion relations at tree level, for Yang-Mills
theory~\cite{Britto:2004ap,Britto:2005fq} and for
gravity~\cite{Bedford:2005yy,Cachazo:2005ca,Benincasa:2007qj}. A
particularly important example is the structure of amplitudes in
$\mathcal{N}=4$ super Yang--Mills theory (SYM), which has remarkable
simplicities obscured by the usual local formulation and
Feynman-diagram calculations. On the other hand, Arkani-Hamed
\emph{et al.} have proposed the idea that $\mathcal{N}=8$
supergravity (SUGRA) may be the quantum field theory with the
simplest amplitudes~\cite{ArkaniHamed:2008gz}, and there is strong
evidence for it: recently there have been intensive studies on both
the hidden symmetries (e.g. $E_{7(7)}$ symmetry,
see~\cite{Kallosh:2008rr, He:2008pb, Brodel:2009hu,Bossard:2010dq}),
and the ultraviolet behavior of the theory
(see~\cite{Green:2006yu,Bern:2007hh,Bern:2008pv,Bern:2009kd,Dixon:2010gz,Elvang:2010jv,Drummond:2010fp,Beisert:2010jx,
Kallosh:2010kk} and references therein).

However, we do not need to go beyond the tree level to see the
simplicity. As shown in~\cite{ArkaniHamed:2008gz}, gravity tree
amplitudes satisfy non-trivial relations, or ``bonus relations",
which are absent in SYM color-ordered amplitudes. These bonus
relations have been applied to MHV amplitudes
in~\cite{Spradlin:2008bu} to show the equivalence of various MHV
formulae in the
literature~\cite{Berends:1988zp,Nair:2005iv,Bedford:2005yy,Elvang:2007sg,Mason:2008jy},
especially to simplify formulae with $(n-2)!$ permutations to those with
$(n-3)!$ permutations. The full strength of these relations, however,
can only be demonstrated when applied to general, non-MHV
amplitudes, and the purpose of the present note is to use bonus
relations to simplify explicit formulae of SUGRA tree amplitudes,
which are obtained by solving BCFW recursion relations. Before
proceeding, let us elaborate on BCFW recursion relations and bonus
relations of SUGRA amplitudes.

Supersymmetric BCFW recursion relations~\cite{Brandhuber:2008pf, ArkaniHamed:2008yf} hold in both SYM and SUGRA
because their amplitudes vanish when two supermomenta are taken to
infinity in a complex
superdirection~\cite{ArkaniHamed:2008yf,ArkaniHamed:2008gz}. More
specifically, under the supersymmetric BCFW shifts of momenta and
$SU(\mathcal{N})$ Grassmannian variables,
\begin{align} \label{shift} \lambda_{\widehat{1}}(z) &= \lambda_1 + z
\lambda_n, \cr \widetilde{\lambda}_{\overline{n}}(z) &=
\widetilde{\lambda}_n - z \widetilde{\lambda}_1, \cr
\eta_{\widehat{n}}(z) &= \eta_n - z \eta_1,
\end{align} SYM and SUGRA amplitudes have at least $1/z$
falloff at large $z$, thus the contour integral $\oint \frac{dz}{z}
M(z)$ can be rewritten as a sum over residues without boundary
contributions, \beq \label{BCFW} M_n=\sum_{L,R} \int
d^{4\mathcal{N}}\eta
M_L(\widehat{1},L,\{-\widehat{P}(z_P),\eta\})\frac{1}{P^2}M_R(\{\widehat{P}(z_P),\eta\},R,\overline{n}),\eeq
where the poles $z=z_P$ are determined by putting the internal
momenta $\widehat{P}(z_P)=\sum_{i\in L}P_i+P_{\widehat{1}}$ on
shell. By solving the recursion relations, explicit formulae for up
to N${}^3$MHV amplitudes, and an algorithm to calculate all tree
amplitudes in SUGRA was proposed in~\cite{Drummond:2009ge}. The
result can be written as a summation over $(n-2)!$ ``ordered gravity
subamplitudes'' with different permutations of particles
$2,\ldots,n-1$. In contrast to SYM color-ordered amplitudes, the
SUGRA amplitudes actually have a faster, $1/z^2$, falloff and the
contour integral $\oint dz M(z)$ gives the bonus relations, \beq
\label{bonus} 0=\sum_{L,R} \int d^8\eta
M_L(\widehat{1},L,\{-\widehat{P}(z_P),\eta\})\frac{z_P}{P^2}M_R(\{\widehat{P}(z_P),\eta\},R,\overline{n}).\eeq
Similar to the MHV case~\cite{Spradlin:2008bu}, we shall see that
these relations can further simplify the explicit formulae for
non-MHV amplitudes by reducing the $(n-2)!$-permutation sum to a new
$(n-3)!$-permutation one.

Another important method that has been widely used to calculate
gravity tree amplitudes are Kawai-Lewellen-Tye (KLT) relations,
first derived in string theory~\cite{Kawai:1985xq} which express
(super)gravity tree amplitudes as sums of products of two copies of
(super)Yang--Mills amplitudes in the field-theory limit. Recently
KLT relations have been proved in
gravity~\cite{BjerrumBohr:2010ta,BjerrumBohr:2010zb} and in
SUGRA~\cite{Feng:2010br} using BCFW recursion relations, without
resorting to string theory. While the well-known KLT relations have
a form of $(n-3)!$ permutations~\cite{Bern:1998sv} (see also
\cite{BjerrumBohr:2010zb}), in the proof it is natural to use the
newly proposed $(n-2)!$ form suitable for BCFW recursion
relations~\cite{BjerrumBohr:2010ta}, and a direct link between these
two forms has been derived in~\cite{Feng:2010hd}. In a related
approach, the so-called square relations between gravity and
Yang-Mills amplitudes, which can be viewed as a reformulation of KLT
relations, have been proposed and proved in~\cite{Bern:2008qj}.
These relations also possess a freedom of going from
$(n-2)!$-permutation form to the simpler $(n-3)!$ form, which,
similar to the freedom in KLT relations, reflects the
Bern-Carrasco-Johansson (BCJ) relations between Yang-Mills
amplitudes~\cite{Bern:2008qj}. For SUGRA amplitudes, the advantage
of having solved BCFW relations to some extent will enable us to go
beyond this implicit freedom following from BCJ relations, and show
the simplification of gravity amplitudes directly in their explicit
forms.

The note is organized as following. In section 2 we briefly review
tree amplitudes in SUGRA and their bonus relations, especially the
simplification of MHV amplitudes when using these relations. Then we
apply these relations to some examples beyond MHV amplitudes,
including the NMHV and N$^2$MHV amplitudes, and prove these
simplified formulae in section 3. The generalization to all tree-level
SUGRA amplitudes are presented in section 4.

\section{A brief review of tree amplitudes in SUGRA and bonus relations}

\subsection{Tree Amplitudes in SUGRA from BCFW Recursion Relations}

By solving Eq. (\ref{BCFW}), all color-ordered SYM tree amplitudes
have been obtained and can be written schematically
as~\cite{Drummond:2008cr}, \beq
A_n(1,...,n)=A^{\rm{MHV}}(1,\ldots,n)\sum_{\alpha}R_{\alpha}(1,\ldots,n),\eeq
where \beq A^{\rm{MHV}}(1,\ldots,n)=\frac{\delta^8(\sum_i
\lambda_i\eta_i)}{\langle 1 2\rangle\langle 2 3 \rangle\cdots\langle
n 1\rangle}\eeq is the MHV superamplitudes, and $R_{\alpha}$ are the
so-called dual superconformal invariants, which, for N${}^k$MHV
amplitudes, are products of $k$ basic invariants of the form,
\begin{equation}
R_{n;a_1b_1;a_2b_2;\ldots;a_rb_r;ab} = \frac{\langle a a-1\rangle
\langle b b-1\rangle \ \delta^{(4)}(\langle \xi | x_{b_r a}x_{ab} |
\theta_{b b_r} \rangle + \langle \xi |x_{b_r b} x_{ba} |
\theta_{ab_r} \rangle)} {x_{ab}^2 \langle \xi | x_{b_r a} x_{ab} | b
\rangle \langle \xi | x_{b_r a} x_{ab} |b-1\rangle \langle \xi |
x_{b_r b} x_{ba} |a\rangle \langle \xi | x_{b_r b} x_{ba}
|a-1\rangle}\,, \label{generalR}
\end{equation}
where the chiral spinor $\xi$ is given by
\begin{equation}
\langle \xi | = \langle n | x_{na_1}x_{a_1 b_1} x_{b_1 a_2} x_{a_2
b_2} \ldots x_{a_r b_r}\,,
\end{equation}
and dual (super)coordinates are defined as
\begin{align} x_{i j} &=
p_i + p_{i+1} + \cdots + p_{j - 1}\,, \cr \theta_{ij} &= \lambda_i
\eta_i + \cdots + \lambda_{j-1} \eta_{j-1}\,.
\end{align}

\begin{figure}
\psfrag{one}[cc][cc]{$1$} \psfrag{a1b1}[cc][cc]{$a_{1}b_{1}$}
\psfrag{a2b2}[cc][cc]{$a_{2}b_{2}$}
\psfrag{a3b3}[cc][cc]{$a_{3}b_{3}$}
\psfrag{b1a1a2b2}[cc][cc]{$a_{1}b_{1};a_{2}b_{2}$}
\psfrag{b2a2a3b3}[cc][cc]{$a_{2}b_{2};a_{3}b_{3}$}
\psfrag{b1a1a3b3}[cc][cc]{$a_{1}b_{1};a_{3}b_{3}$}
\psfrag{b1a1b2a2a3b3}[cc][cc]{$a_{1}b_{1};a_{2}b_{2};a_{3}b_{3}$}
\psfrag{two}[cc][cc]{$2$} \psfrag{n1}[cc][cc]{$n$}
\psfrag{a1p}[cc][cc]{$a_{1}$} \psfrag{a2p}[cc][cc]{$a_{2}$}
\psfrag{b1}[cc][cc]{$b_{1}$} \psfrag{b2}[cc][cc]{$b_{2}$}
\centerline{{\epsfysize6cm \epsfbox{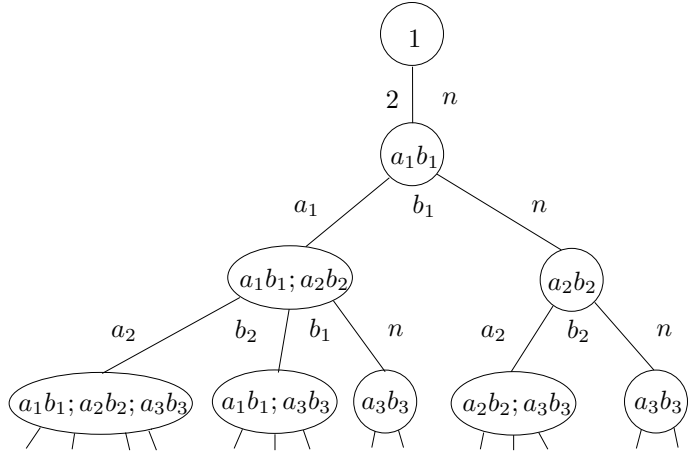}}} \caption{An rooted
tree diagram for tree-level SYM amplitudes. The figure is the same
as the tree diagram presented in \cite{Drummond:2009ge}.}
\label{newYMtree}
\end{figure}

\begin{figure}
\psfrag{uvs}[cc][cc]{$u_{1}v_{1};\ldots u_{r}v_{r};a_{p-1}b_{p-1}$}
\psfrag{uvab}[cc][cc]{$u_{1}v_{1};\ldots
u_{r}v_{r};a_{p-1}b_{p-1};a_{p}b_{p}$}
\psfrag{uvnext}[cc][cc]{$u_{1}v_{1};\ldots u_{r}v_{r};a_{p}b_{p}$}
\psfrag{ab}[cc][cc]{$a_{p}b_{p}$} \psfrag{up}[cc][cc]{$a_{p-1}$}
\psfrag{bp1}[cc][cc]{$b_{p-1}$} \psfrag{bvr}[cc][cc]{$v_{r}$}
\psfrag{bv1}[cc][cc]{$v_{1}$} \psfrag{n1}[cc][cc]{$n$}
\psfrag{dots}[cc][cc]{$\ldots$} \centerline{{\epsfysize3cm
\epsfbox{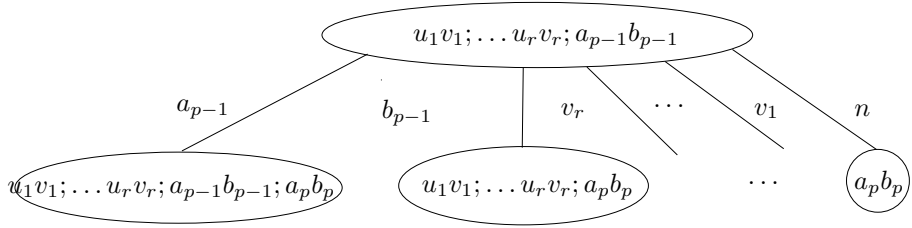}}}  \caption[]{ The rule for going from
line $p-1$ to line $p$ (for $p>1$) in~\fig{newYMtree}. For every
vertex in line $p-1$ of the form given at the top of the diagram,
there are $r+2$ vertices in the lower line (line $p$). The labels in
these vertices start with $u_{1}v_{1};\ldots
u_{r}v_{r};a_{p-1}b_{p-1};a_{p}b_{p}$ and they get sequentially
shorter, with each step to the right removing the pair of labels
adjacent to the last pair $a_p,b_p$ until only the last pair is
left. The summation limits between each line are also derived from
the labels of the vertex above. The left superscripts which appear
on the associated $R$-invariants start with $u_1v_1\ldots u_r v_r
b_{p-1} a_{p-1}$ for the left-most vertex. The next vertex to the
right has the superscript $u_1v_1 \ldots u_r v_r a_{p-1} b_{p-1}$,
i.e. the same as the first but with the final pair in alphabetical
order. The next vertex has the superscript $u_1v_1 \dots u_r v_r$
and thereafter the pairs are sequentially deleted from the right.}
\label{cluster}
\end{figure}

There is only one invariant $R=1$ for MHV case, while we have a sum
of $R_{n;a_1 b_1}$ with $1<a_1<b_1<n$ for NMHV case. Furthermore,
for N${}^2$MHV case we have $R_{n;a_1 b_1}R^{b_1 a_1}_{n;a_1 b_1,a_2
b_2}$ with $1<a_1<a_2<b_2 \leq b_1<n$ and $R_{n;a_1 b_1}R^{a_1
b_1}_{n;a_2 b_2}$ with $1<a_1<b_1\leq a_2<b_2<n$, where superscripts
denote boundary modifications of these
invariants~\cite{Drummond:2008cr}.

Generally the summation variables $\alpha$, and boundary
modifications, can be represented by a rooted
tree diagram~\cite{Drummond:2008cr,Drummond:2009ge} (see
Fig.~\ref{newYMtree} and Fig.~\ref{cluster}). For N$^k$MHV
amplitudes, there are $C_k=\frac{(2k)!}{k!(k+1)!}$ (Catalan number)
types of terms labeled by $\alpha$'s corresponding to a path from
the root to the $k$-th level in Fig.~\ref{newYMtree}, and each type
can be written as a list of $k$ pairs of labels with a particular
order between them, $\alpha \equiv \{n;a_1,b_1;\ldots;a_k,b_k\}$. Not only
does the summation over $\alpha$ include all types of terms, but it
also sums over all possible $1<a_i,b_i<n$ in the corresponding
order.

In~\cite{Drummond:2009ge}, solving Eq.~(\ref{BCFW}) for SUGRA is
simplified by using ordered gravity subamplitude $M(1,\ldots,n)$,
which satisfy the ordered BCFW recursion relations similar to
Yang-Mills theory,
\begin{equation} \label{mdef} M(1,\ldots,n) \equiv \sum_{i=3}^{n-1}
\int \frac{d^8\eta}{P^2} M(\widehat{1},2,\ldots,i-1,\widehat{P})
M(-\widehat{P},i,\ldots,n-1,\overline{n})\,,
\end{equation}
and the sum of $(n-2)!$ permutations of ordered gravity
subamplitudes gives the full amplitude, \beq\label{n-2form}
\mathcal{M}_n=\sum_{\mathcal{P}(2,3,\ldots,n-1)} M(1,\ldots,n).\eeq

A solution for $M(1,\ldots,n)$ is obtained
in~\cite{Drummond:2009ge}, \beq
M(1,...,n)=[A^{\rm{MHV}}(1,\ldots,n)]^2\sum_{\alpha}G_{\alpha}R^2_{\alpha}(1,\ldots,n),
\eeq where the invariants $R_{\alpha}$ are exactly the same as those
in SYM (including boundary modifications), namely products of basic
invariants (\ref{generalR}), with the same set of summation
variables $\alpha$ as given in Fig.~\ref{newYMtree} and
Fig.~\ref{cluster}, and the `dressing factors', $G_{\alpha}$, are
independent of the Grassmannian variables $\eta_i$, and they break
dual conformal invariance of the SYM solution. These factors have
been calculated explicitly for up to N${}^3$MHV amplitudes, for
example MHV case,
\begin{equation}
\label{MHVG} G^{\rm MHV}(1,\ldots,n) = x_{13}^2 \prod_{s=2}^{n-3}
\frac{\langle s | x_{s,s+2} x_{s+2,n} |n\rangle}{\langle s
n\rangle}\,,
\end{equation}
and there is an algorithm to calculate them in general cases, but we
do not need their expressions in this note. In addition, tree-level
amplitudes of $n$-graviton scattering can be obtained from SUGRA
superamplitudes (\ref{n-2form}), by choosing fermionic coordinates
$\eta=0$ for positive-helicity gravitons, and integrating over
$d^8\eta$ for negative-helicity ones. Details of the solution can be
found in~\cite{Drummond:2009ge}.

Therefore, SUGRA tree amplitude can be written as a summation of
$(n-2)!$ ordered gravity subamplitudes, and each of them has a
structure similar to SYM ordered amplitude. In the following we
shall use bonus relations to reduce this form to a simpler, $(n-3)!$
form, and first we recall the simplest MHV case.

\subsection{Applying Bonus Relations to MHV Amplitudes}

Applying bonus relation to MHV SUGRA tree-level amplitudes was well
understood in \cite{Spradlin:2008bu}. From Eq.~(\ref{MHVG}), we have
the MHV amplitudes as a summation of $(n-2)!$ terms, \beq
\mathcal{M}_n^{\rm{MHV}}=G^{\rm MHV}(1,\ldots
n)[A^{\rm{MHV}}(1,\ldots,n)]^2+\mathcal{P}(2,3,\ldots,n-1).\eeq

\begin{figure}[t]
  \begin{center}
\includegraphics[scale=0.75]{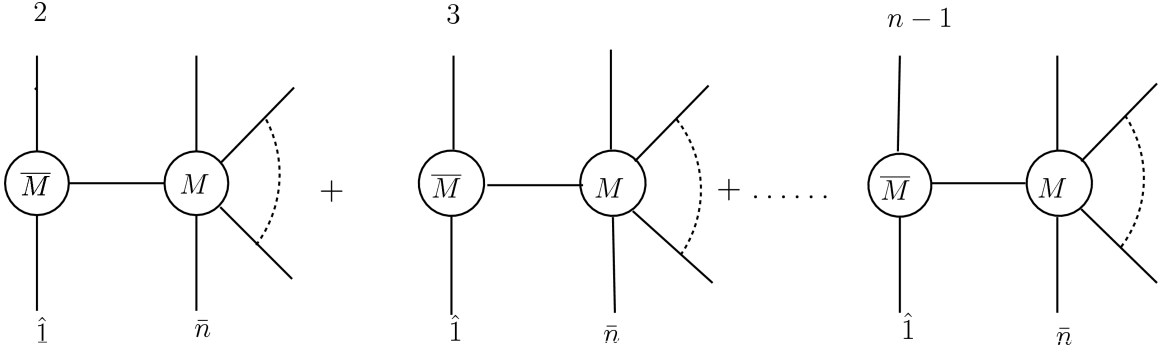}
\end{center}
  \caption{All factorizations contributing to (\ref{MHV-chanel}) for the MHV amplitude.}
  \label{mhvfig}
\end{figure}

From Fig.~\ref{mhvfig}, we see that there are $(n-2)$ BCFW
factorizations and thus the formula can be expressed as, \beq
\label{MHV-chanel}
\mathcal{M}_n^{\rm{MHV}}=M_2+M_3+\ldots+M_{n-1},\eeq where each
$M_i$ is a BCFW term from
$\overline{\rm{MHV}}(\widehat{1},i,\widehat{P}(z_i))\times
\rm{MHV}_{n-1}$ with $z_i=-\frac{\langle 1 i\rangle}{\langle n
i\rangle}$. Now since the amplitude has $1/z^2$ fall off, we have a
bonus relation which is simple in the MHV case, \beq
0=z_2M_2+z_3M_3+\ldots+z_{n-1}M_{n-1}.\eeq  Using this relation, we
can express the last diagram $M_{n-1}$ in terms of the other $n-3$
diagrams, and a simple manipulation gives us a $(n-3)!$-term
formula,
\begin{equation}
\begin{aligned} \label{MHV-n}
\mathcal{M}_n^{\rm MHV}&=B^{\rm MHV} G^{\rm MHV}(1,2,\dots,n) [A^{\rm MHV}(1,2,\dots,n)]^2 \\
&+ \mathcal{P}(2,3,\dots,n-2).
\end{aligned}
\end{equation}
where we have defined the MHV bonus coefficient $ B^{\rm MHV}=\frac
{\langle 1~n \rangle \langle n-1~n-2 \rangle }{ \langle 1~n-1
\rangle \langle n~ n-2 \rangle}$. Beyond MHV, we have many more types
of BCFW diagrams with complicated structures and the application of bonus relations becomes
trickier. In the next section, we shall work out the NMHV and
N${}^2$MHV cases, and then move on to general amplitudes in section 4.

\section{Applying Bonus Relations to Non-MHV Gravity Tree Amplitudes}

\subsection{General Strategy}

\begin{figure}[t]
  \begin{center}
  \vspace{-0.9cm}
   \hspace{-1.cm} \subfigure[ Inhomogeneous diagram type I]{\label{NkMHV-1}\includegraphics[scale=0.7]{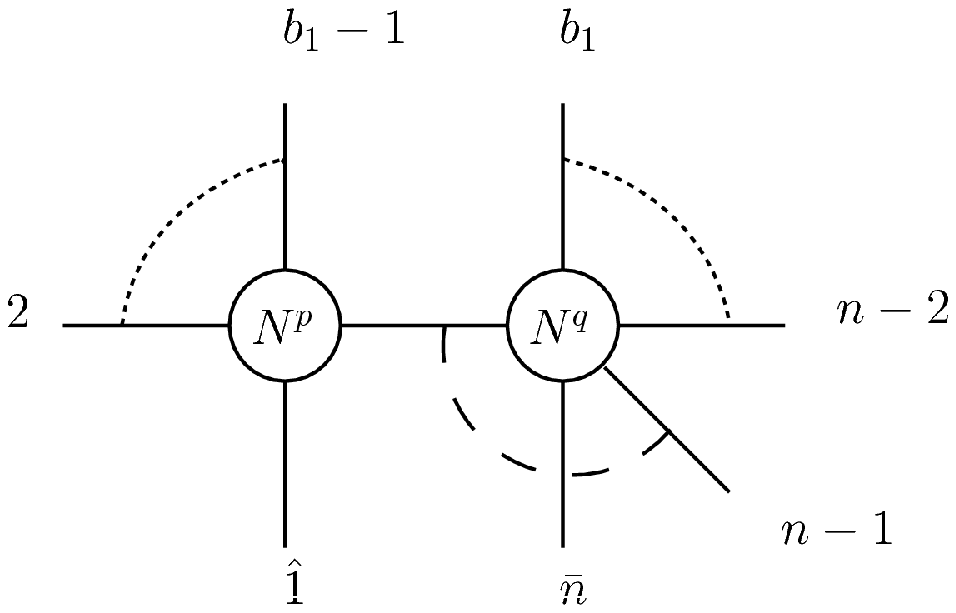}}
      \hspace{2.cm} \subfigure[ Inhomogeneous diagram type II]{\label{NkMHV-2}\includegraphics[scale=0.7]{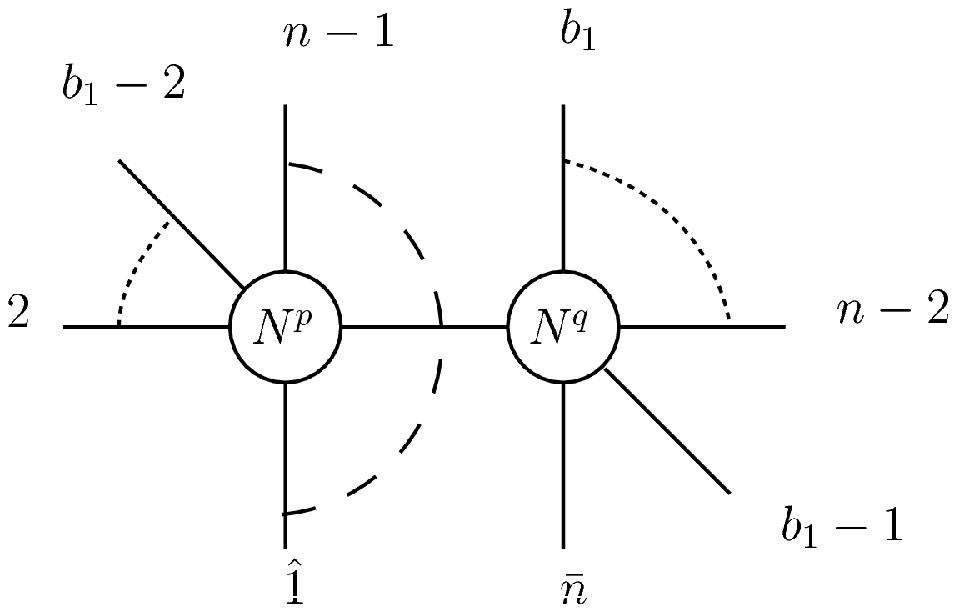}}
       \subfigure[ Homogeneous diagram]{\label{NkMHV-3}\includegraphics[scale=0.7]{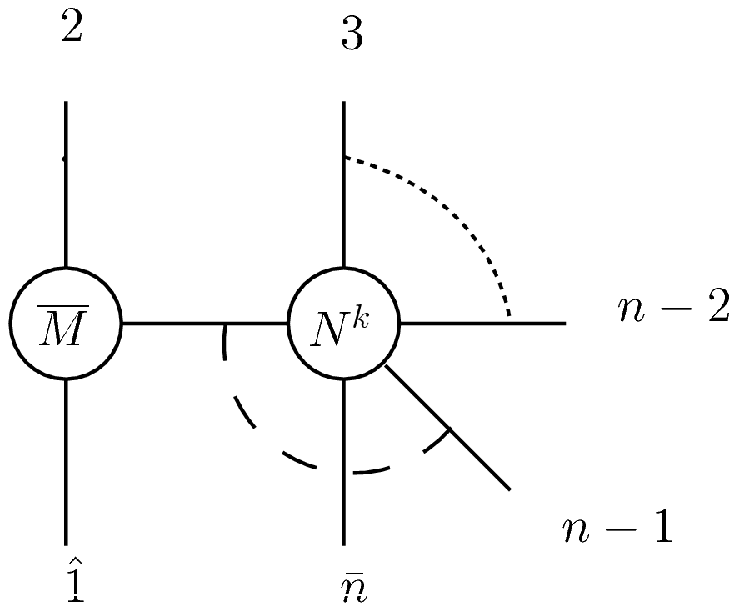}}
    \hspace{3cm}\subfigure[ Diagram deleted by bonus relations]{\label{NkMHV-delete}\includegraphics[scale=0.7]{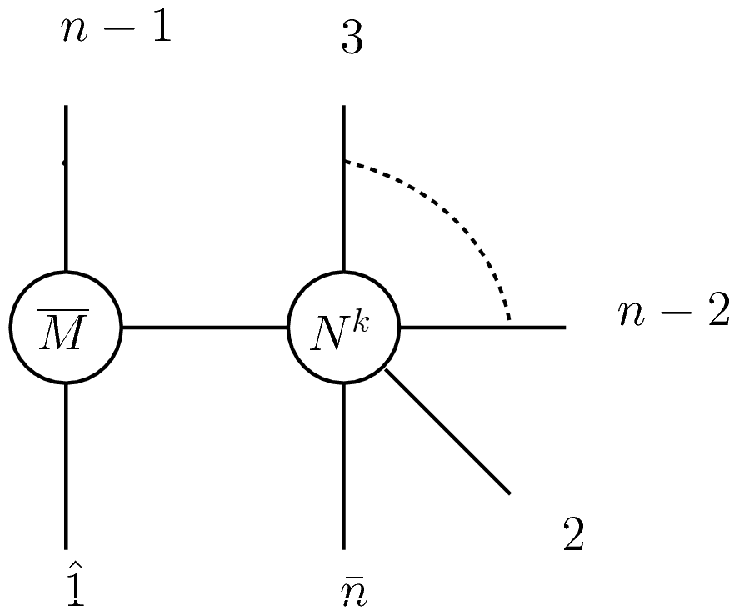}}
  \end{center}
  \caption{ Different types of diagrams for a general N$^{k}$MHV amplitude, where $k=p+q+1$.
We use a dashed line $----$ connecting three legs to denote a
bonus-simplified lower-point amplitude, in which these three legs are
kept fixed. For lower-point amplitudes without dashed lines, we use
the usual $(n-2)$! form.}
  \label{N^k}
\end{figure}

Before moving on to examples, we first explain the general strategy for
applying bonus relations to non-MHV gravity tree amplitudes. For a
N$^k$MHV amplitude, inhomogeneous contributions of the form N$^p$MHV
$\times$ N$^q$MHV are needed $(p+q+1=k)$\footnote{We follow the
notations of reference \cite{Drummond:2008cr} to call the
contributions from diagrams of type  Fig.~\ref{NkMHV-1} or
Fig.~\ref{NkMHV-2} as inhomogeneous contributions, while those from
Fig.~\ref{NkMHV-3} as homogeneous ones.}. Naively one would like to
use ``bonus-simplified"\footnote{Here ``bonus-simplified" means that
these lower-point amplitudes used in the BCFW diagrams are
simplified by using bonus relations.} lower-point amplitudes for both
$M_L$ and $M_R$ in Eq.~(\ref{BCFW}), but this is not compatible with
the fact that we can only delete one diagram (not two) by applying
the bonus relations (\ref{bonus}), if we want to preserve the
structure of ordered BCFW recursion relations.

To keep the advantages of the ordered BCFW recursion relations,
which are crucial to solve for all tree-level amplitudes, instead we
shall apply bonus relations selectively. The idea is
illustrated in Fig.~\ref{N^k}. Similar to the MHV case, we shall
delete Fig.~\ref{NkMHV-delete} by using bonus relations
(\ref{bonus}). To compute the inhomogeneous parts of the amplitudes,
we shall use the bonus-simplified amplitude only on one side of a
BCFW diagram, namely the lower-point amplitude with the leg $(n-1)$ in it,
as indicated in Fig.~\ref{NkMHV-1} and Fig.~\ref{NkMHV-2}.
In this way, the amplitude splits into two types, one type coming
from the diagrams of the form as in Fig.~\ref{NkMHV-1}, which has the leg
$(n-1)$ adjacent to the leg $n$ and will be called the normal, or type I
contributions, and the other one coming from those having the form
as in Fig.~\ref{NkMHV-2}, which has the leg $(n-1)$ exchanged with another
leg $(b_1-1)$, and will be called the exchanged, or type II contributions,
\begin{equation}
\begin{aligned}
\mathcal{M}_n=\big[A_n^{\rm MHV} \big]^2 \big(\sum_{\alpha}
B^{(1, m_1)}_{\alpha} G_{\alpha} R^2_{\alpha} + \sum_{\beta}
B^{(2, m_2)}_{\beta} [G_{\beta} R^2_{\beta}(b_1-1 \leftrightarrow n-1)]
\big) + \mathcal{P}(2,3,\dots,n-2),
\end{aligned}
\end{equation}
where $(b_1-1 \leftrightarrow n-1)$ denotes the exchanges of momenta
$(p_{b_1-1} \leftrightarrow p_{n-1})$ as well as the fermionic
coordinates $(\eta_{b_1-1} \leftrightarrow \eta_{n-1})$, and we have used square bracket to indicate that
the exchanges act only on the expression inside the bracket. The
superscript $(i,m_i)$ in $B^{(i,m_i)}_{\alpha}$ is used to show the type of this
contribution, which will become clear in the examples.

Thus we have seen that, by using bonus relations, any amplitude can
be written as a summation of $(n-3)!$ permutations with the coefficients
$B^{(i,m_i)}_{\alpha}$, which will be called bonus coefficients. In this
section, we shall calculate all bonus coefficients for NMHV and
N$^2$MHV cases, and generalize the pattern observed in these
examples to general N$^k$MHV amplitudes in the next section. Once
bonus coefficients are calculated, we obtain explicitly all
simplified SUGRA tree amplitudes.

\subsection{NMHV Amplitudes}

Here we use bonus relations to simplify the $(n-2)!$ form of NMHV
amplitudes. First we shall state the general simplified form of NMHV
amplitudes, and then prove it by induction. To be concise, we
abbreviate the combinations
\begin{equation}
\begin{aligned}
\{n;a_1b_1\} \equiv G_{n;a_1b_1} \big[ R_{n;a_1b_1}A^{{\rm MHV}}(1,2,\dots, n) \big]^2
\end{aligned}
\end{equation}
and similar notations will be used in the following sections.
\begin{figure}
  \begin{center}
  \hspace{-1.cm}
\subfigure[ Inhomogeneous diagram type
I]{\label{NMHV-1}\includegraphics[scale=0.6]{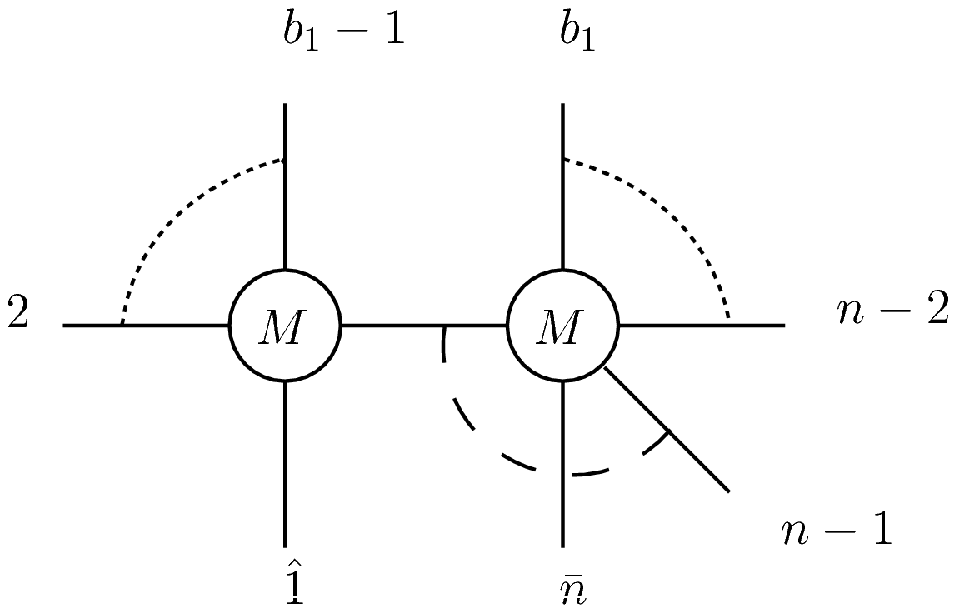}}
\subfigure[ Inhomogeneous diagram type
II]{\label{NMHV-2}\includegraphics[scale=0.6]{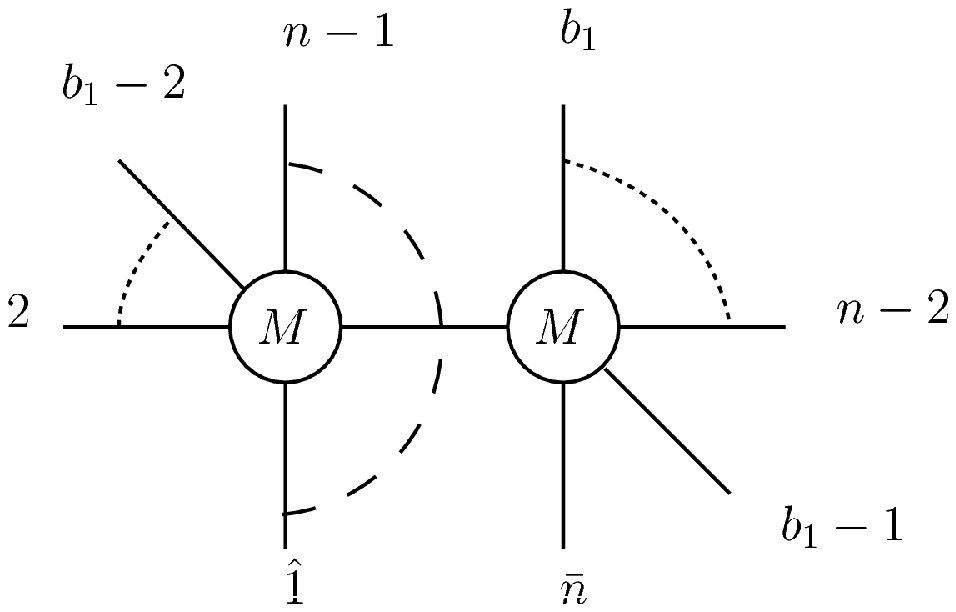}}
\subfigure[ Homogeneous
diagram]{\label{NMHV-hom-1}\includegraphics[scale=0.6]{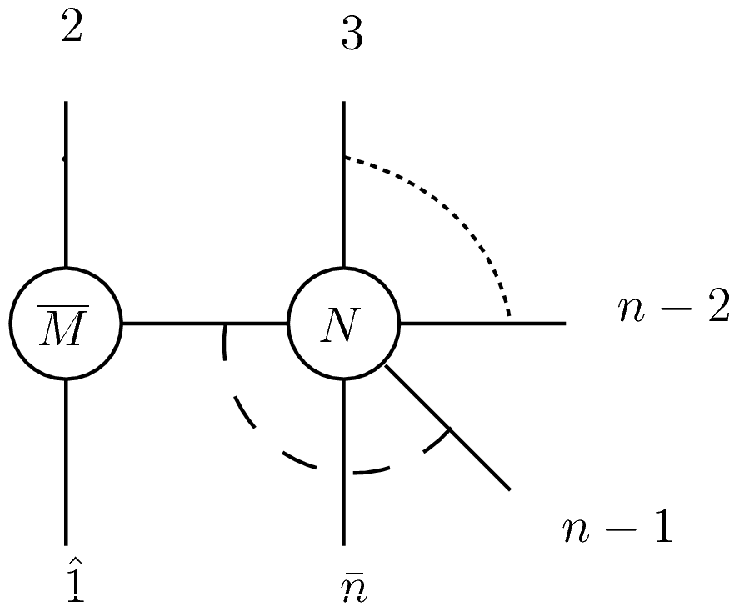}}
     \end{center}
  \caption{ Diagrams for NMHV amplitudes.}
  \label{NMHV}
\end{figure}

As mentioned above generally, we delete the contributions
corresponding to Fig.~\ref{NkMHV-delete} by using the bonus relation
(\ref{bonus}). It is straightforward to compute the inhomogeneous
contributions from the two MHV $\times$ MHV diagrams, Fig.~\ref{NMHV-1}
and Fig.~\ref{NMHV-2}. Firstly, let us consider the contribution from
Fig.~\ref{NMHV-1}, which corresponds to terms with $a_1=2$, and we have
\begin{equation}
\begin{aligned}
M_1=B^{(1)}_{n;2b_1} \{n;2 b_1\},~~~ {\rm with} ~~ 4 \leq b_1 \leq
n-1,
\end{aligned}
\end{equation}
where $B^{(1)}_{n;2b_1}$ are the special cases of the general bonus
coefficients $B^{(1)}_{n;a_1b_1}$. We have used the superscript $(1)$ to indicate that
this is the contribution coming from type-I diagram, and similar notations will be used below.

When $b_1 \neq n-1$, the bonus coefficients are given by,
\begin{equation}
\begin{aligned} \label{B1-1}
B^{(1)}_{n;a_1b_1}= B^{\rm MHV} \frac {\langle n-1| x_{b_1a_1} x_{b_1n} |n \rangle }{ \langle n-1|x_{b_1a_1}x_{a_1n}|n \rangle}.
\end{aligned}
\end{equation}
Here we note that we can get the above coefficients from the previous ones,
namely the bonus coefficients of MHV amplitude, multiplied by the
factor $\frac {\langle n-1| x_{b_1a_1} x_{b_1n} |n \rangle }{
\langle n-1|x_{b_1a_1}x_{a_1n}|n \rangle}.$
It is a general feature
of this type of coefficients, for N$^k$MHV case they are given by
N$^{k-1}$MHV coefficients multiplied by the same factor, as we shall
see explicitly again in the N$^2$MHV case.

However when $b_1=n-1$, no bonus relation can be used for the
right-hand-side 3-point MHV amplitude in Fig.~\ref{NMHV-1}, and we
find
\begin{equation}
\begin{aligned} \label{B1-2}
B^{(1)}_{n;a_1 n-1}=\frac {\langle 1~n \rangle  }{  \langle 1~n-1 \rangle} \frac {\langle n-1 |x_{n-1a_1} |n-1 ] }{
\la n|x_{na_1}|n-1]}.
\end{aligned}
\end{equation}
For the exchanged diagrams, Fig.~\ref{NMHV-2}, the contribution can
be similarly obtained
\begin{equation}
\begin{aligned}
M_2 &= B^{(2)}_{n;2b_1} [\{n;2 a_1\}(b_1-1 \leftrightarrow n-1)],~~~ {\rm with} ~~4 \leq b_1 \leq n-1,
\end{aligned}
\end{equation}
where the bonus coefficients $B^{(2)}_{n;a_1b_1}$ are given by
\begin{equation}
\begin{aligned} \label{B2}
B^{(2)}_{n;a_1b_1}&=\frac {\langle 1~n \rangle }{ \langle 1~n-1 \rangle} \frac {\langle n-1~b_1-2 \rangle (x'_{a_1b_1})^2  }{ \langle n|x_{na_1}x'_{a_1b_1}|b_1-2 \rangle},
\end{aligned}
\end{equation}
and we have defined $x'_{a_ib_i}$ as,
\beqa  x'_{a_ib_i} &\equiv &x_{a_ib_i-1}+x_{n-1n} \nn
                   &= &x_{a_ib_i}(p_{b_i-1} \leftrightarrow p_{n-1}).
\eeqa

\begin{figure}
  \begin{center}
\hspace{-1.cm}\subfigure[ $5$-point diagram deleted by bonus
relation]{\label{NMHV-special-delete}\includegraphics[scale=0.7]{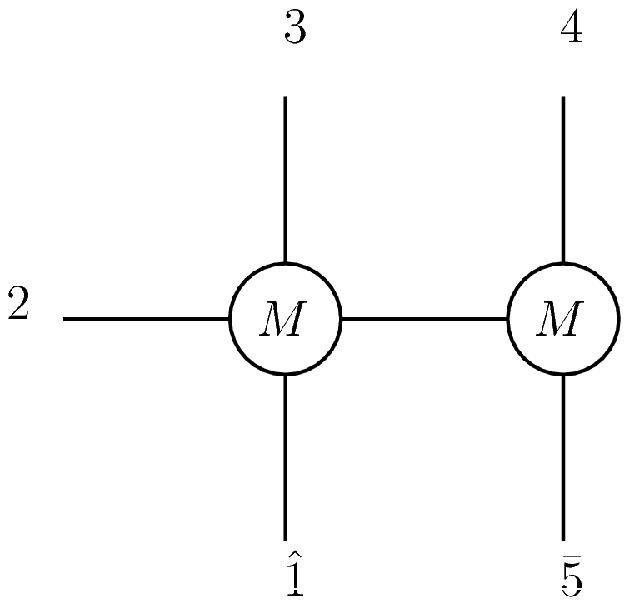}}
   \subfigure[ $5$-point diagram]{\label{NMHV-special-1}\includegraphics[scale=0.7]{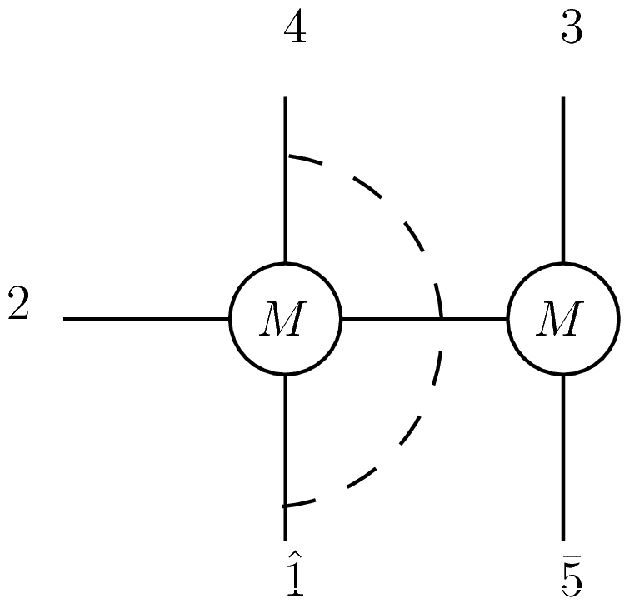}}
   \hspace{1.cm}\subfigure[ $6$-point diagram calculating the boundary contribution]{\label{NMHV-special-2}\includegraphics[scale=0.7]{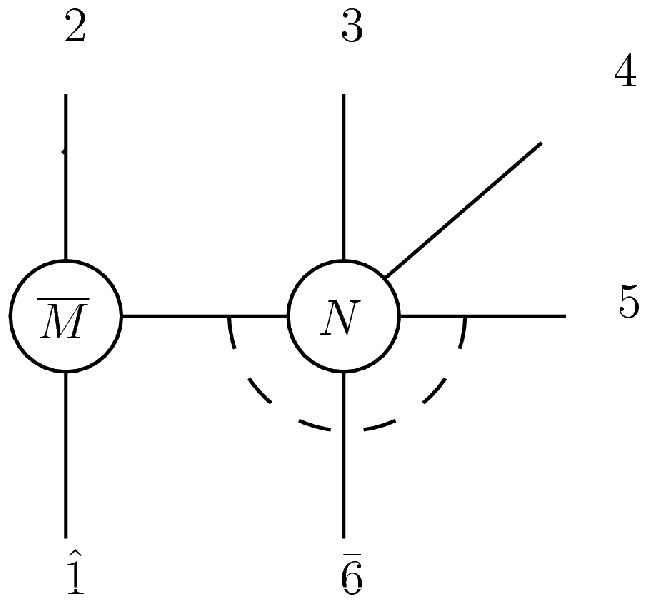}}
  \end{center}
 \caption{ Diagrams for $5$-point NMHV amplitude and the boundary term of $6$-point NMHV amplitude.
 Fig.~6(a) and Fig.~6(b) are used to calculate the bonus-simplified,
$5$-point, right-hand-side amplitude of Fig.~6(c). }
  \label{NMHV-others}
\end{figure}

All the above calculations do not include the boundary case
$a_1=n-3, b_1=n-1$, which needs special treatment. This boundary
case is special because it recursively reduces to the special 5-point NMHV
($\overline{\rm{MHV}}$) amplitude. It does not have the diagram of
the type $\overline{{\rm MHV}}_3 \times$ NMHV, and one has to
treat it separately. We apply the bonus relations to this case in
the following way: we use Eq.~(\ref{bonus}) to delete the
contribution from Fig.~\ref{NMHV-special-delete}, and compute
Fig.~\ref{NMHV-special-1}, and we find
\begin{equation}
\begin{aligned}\label{5pt-boundary}
\mathcal{M}_5=-\frac {[24][34][51] }{ [23][45][41]} \big[ \{5;24\} (3
\leftrightarrow 4) \big] + \mathcal{P}(2,3).
\end{aligned}
\end{equation}
By plugging the above 5-point result in Fig.~\ref{NMHV-special-2}, we get the boundary
term of the 6-point NMHV amplitude
\begin{equation}
\begin{aligned} \label{6pt-boundary}
M^{\rm (boundary)}_6=\frac {\langle 16 \rangle \langle 2 5  \rangle [35][45]
x_{36}^2 }{ \langle 15 \rangle[34] \langle 2|1+6|5]\langle
6|1+2|5]} \big[ \{6;35\} (4 \leftrightarrow 5) \big].
\end{aligned}
\end{equation}
A generic form for the boundary term of the $n$-point NMHV
amplitudes can be obtained as a straightforward generalization of
(\ref{5pt-boundary}) and (\ref{6pt-boundary}),
\begin{equation} \label{NMHV-boundary}
\begin{aligned}
M^{(\rm{boundary})}_n=B^{({\rm boundary})}_{n;n-3~n-1} \big[ \{n;n-3~n-1\} (n-2
\leftrightarrow n-1) \big],
\end{aligned}
\end{equation}
where $B^{(\rm boundary)}_{n;n-3~n-1}$ is given by,
\begin{equation}
\begin{aligned} \label{BNHMV-boundary}
B^{({\rm boundary})}_{n;n-3~n-1}= \frac {\langle 1n \rangle \langle n-4~n-1  \rangle [n-3~n-1][n-2~n-1] x^2_{n-3n} }{ \langle 1~n-1 \rangle [n-3~n-2]\langle n-4|x_{n-3~n-1}|n-1]\langle n|x_{n-1~n-3}|n-1] }.
\end{aligned}
\end{equation}

Putting everything together, we obtain the general formula for NMHV
amplitude and as promised, the amplitude indeed can be written as a sum of $(n-3)!$ permutations
\begin{eqnarray}
\mathcal{M}_n^{\rm{NMHV}}&=&\sum^{n-4}_{a_1=2} \sum^{n-1}_{b_1=a_1+2}
\left( B^{(1)}_{n;a_1b_1} \{n;a_1b_1\} + B^{(2)}_{n;a_1b_1} [\{n;a_1b_1\} (b_1-1
\leftrightarrow n-1)] \right) + M^{(\rm{boundary})}_n  \cr
 &&+\mathcal{P}(2,3,\dots,n-2).
 \label{NMHVfinal}
\end{eqnarray}

\subsubsection{\bf Proof by Induction}

Here we shall give an inductive proof for the simplified NMHV formula.
For $a_1=2$, as we explained above, the formula follows directly
from Fig.~\ref{NMHV-1} and Fig.~\ref{NMHV-2}. Therefore we shall
focus on the cases when $a_1 \geq 3$, which correspond to the
homogeneous contributions from Fig.~\ref{NMHV-hom-1}. We shall prove
that the formula satisfies the BCFW recursion relations.

First note that we have deleted one diagram of the form
$M^{\rm{MHV}}_L(\hat{1},n-1,\hat{P})\times M^{\rm{MHV}}_R$ by using
bonus relations, this results in a multiplicative prefactor for the
overall amplitude, which is given by, \beq
 (1-\frac{z_{2}}{z_{n-1}})=\frac{\langle
1n\rangle \langle n-1~2\rangle}{\langle n2\rangle \langle 1
n-1\rangle}. \label{prefactor} \eeq

Let us consider the bonus coefficient $B^{(1)}_{n;a_1 b_1}$, other coefficients $B^{(2)}_{n;a_1 b_1}$
and $B^{({\rm boundary})}_{n;n-3~n-1}$ can be treated similarly.
By plugging formula (\ref{B1-1}) into the $(n-1)$-point amplitude
$M(-{\hat P}, 3, 4, \ldots,n-1, {\bar n})$ in Fig.~\ref{NMHV-hom-1}, it
is straightforward to check that the second piece of $B^{(1)}_{n;a_1
b_1}$, $\frac {\langle n-1| x_{b_1a_1} x_{b_1n} |n \rangle }{ \langle n-1|x_{b_1a_1}x_{a_1n}|n \rangle}$,
is transformed back to itself under the recursion relations.

For the first piece $B^{\rm MHV}=\frac {\langle n-1~n-2 \rangle
\langle 1~n \rangle }{  \langle n~n-2 \rangle \langle 1~n-1
\rangle}$ of $B^{(1)}_{n;a_1 b_1}$, which is the MHV bonus coefficient, the proof is
essentially the same as in the MHV case. Taking into account the
factor in (\ref{prefactor}) coming from bonus relations, we have
\begin{equation}
\begin{aligned}
\frac {\langle n-1~n-2 \rangle \langle {\hat p}~n \rangle }{  \langle n~n-2 \rangle \langle {\hat p}~n-1 \rangle}
\times \frac {\langle 1~n \rangle \langle n-1~2 \rangle }{  \langle 1~n-1 \rangle \langle n~2 \rangle}
=\frac {\langle n-1~n-2 \rangle \langle 1~n \rangle }{  \langle n~n-2 \rangle \langle 1~n-1 \rangle}.
\end{aligned}
\end{equation}
Thus the contribution with $B^{(1)}_{n;a_1b_1}$ indeed satisfies the
recursion relations.

A final remark is in order. We have used in the proof that
$\{n;a_1b_1\}$ satisfy the ordered BCFW recursion relations by
themselves.

\subsection{N${}^2$MHV amplitudes}

In this subsection we consider N${}^2$MHV amplitudes as one more
example to show the general features of bonus-simplified gravity
amplitudes. Similar to NMHV case, let us denote the ordered gravity
solutions in the following way
\begin{equation}
\begin{aligned}
H^{(1)}_{n;a_1b_1,a_2b_2} \big[ R_{n;a_1b_1}R^{b_1a_1}_{n;a_1b_1,a_2b_2}A^{{\rm
MHV}}(1,2,\dots, n) \big]^2 &\equiv& \{ n;a_1b_1,a_2b_2\}_1, \nn
 H^{(2)}_{n;a_1b_1,a_2b_2} \big[ R_{n;a_1b_1}R^{a_1b_1}_{n;a_2b_2}A^{{\rm
MHV}}(1,2,\dots, n) \big]^2 &\equiv&  \{ n;a_1b_1,a_2b_2 \}_2.
\end{aligned}
\end{equation}

\begin{figure}
  \begin{center}

  \hspace{-1.8cm} \subfigure[ Homogeneous diagram]{\label{NNMHV-special-1}\includegraphics[scale=0.63]{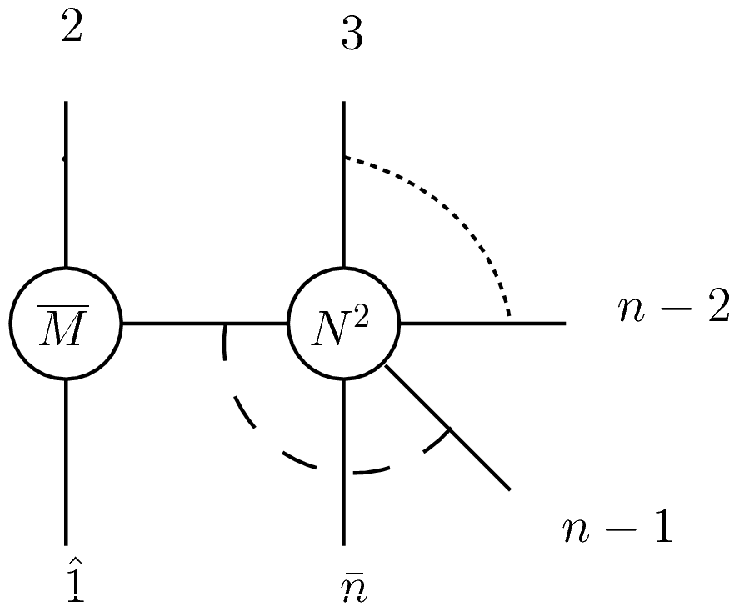}}
 \hspace{0.3cm}\subfigure[ Inhomogeneous diagram type~I]{\label{NNMHV-1}\includegraphics[scale=0.63]{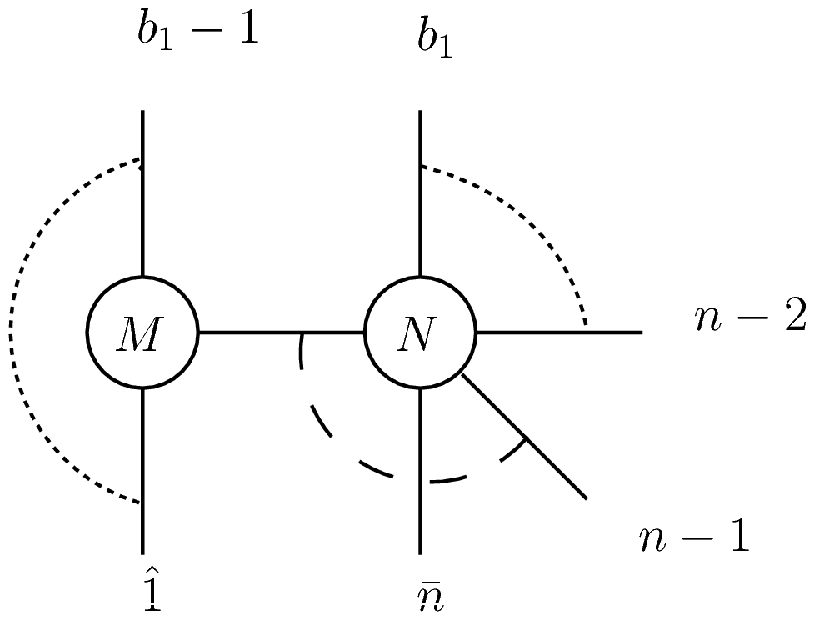}}
    \hspace{0.7cm}\subfigure[ Inhomogeneous diagram type~II]{\label{NNMHV-4}\includegraphics[scale=0.63]{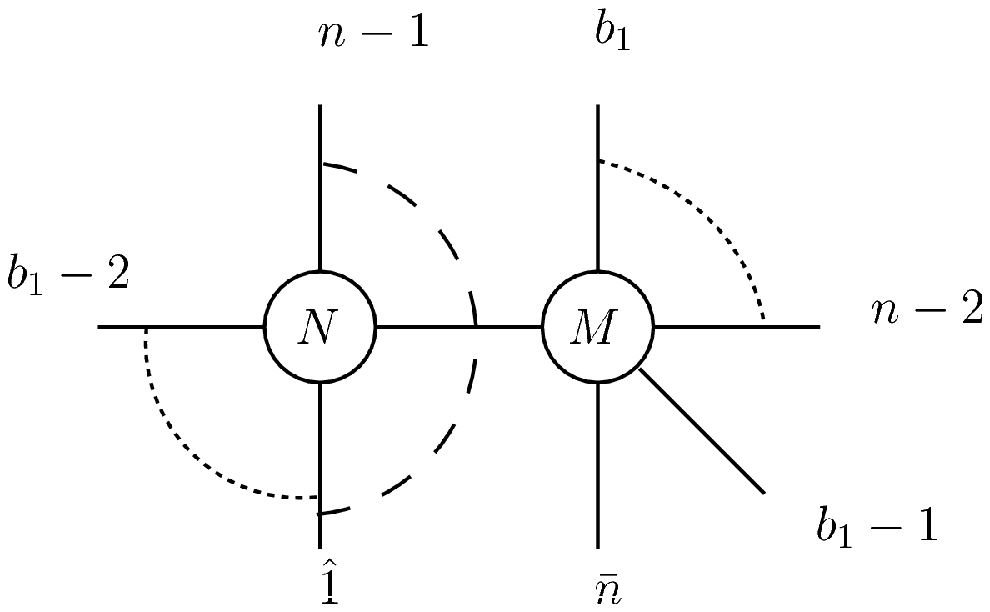}}

  \hspace{-1.cm}
    \subfigure[ Inhomogeneous diagram type~II]{\label{NNMHV-3}\includegraphics[scale=0.7]{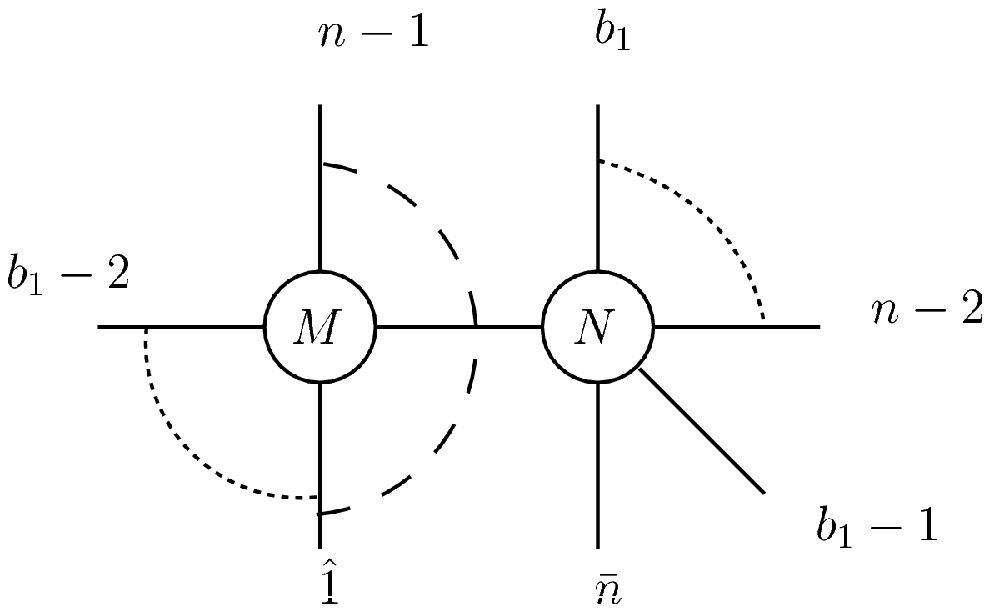}}
    \hspace{1.cm}\subfigure[ Inhomogeneous diagram type~I]{\label{NNMHV-2}\includegraphics[scale=0.66]{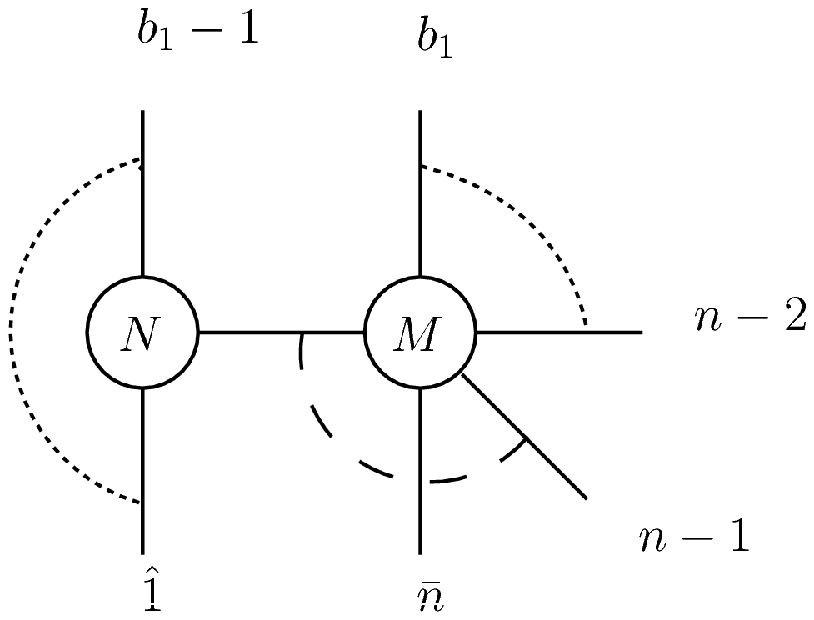}}

  \end{center}
  \caption{ Diagrams for N${}^2$MHV amplitudes.}
  \label{NNMHV}
\end{figure}

There are four relevant types of diagrams (and a boundary case)
which contribute to the general N$^2$MHV amplitudes. The general
structure of N$^2$MHV is given in Fig.~\ref{NNMHV} and the
corresponding contributions from each of the four diagrams can be
calculated separately.

First we consider the contributions from the diagrams in
Fig.~\ref{NNMHV-1}, which are of the form MHV$\times$ NMHV. We use
bonus-simplified amplitude for the right-hand-side NMHV amplitude
and we obtain\footnote{Here and in the following calculations we
have included the corresponding homogeneous terms, for the case we
consider the contributions are from Fig.~\ref{NNMHV-special-1}.},
\beqa M_{{\rm I}}&=&\sum_{2\leq a_1, b_1\leq n-1}\sum_{b_1\leq a_2,
b_2<n} \big( B_{n;a_1b_1;a_2b_2}^{(1,1)}\{n;a_1b_1;a_2b_2\}_2 \nn
&~&+ ~
B_{n;a_1b_1;a_2b_2}^{(1,2)}[\{n;a_1b_1;a_2b_2\}_2(b_2-1\leftrightarrow
n-1)] \big ) \nn &~&+  \sum_{2\leq a_1, b_1\leq n-1}
B_{n;a_1b_1;n-3n-1}^{(1,{\rm
boundary})}[\{n;a_1b_1;n-3n-1\}_2(n-2\leftrightarrow n-1)], \eeqa
where in the first sum $a_2 \leq n-4$ because of the range of
summation of the first term in Eq.~(\ref{NMHVfinal}). Here the bonus
coefficients are given by \beqa \label{N^2MHV-1}
B_{n;a_1b_1;a_2b_2}^{(1,1)}&=&\frac{\langle 1n\rangle \langle
n-1~n-2\rangle  \langle n-1|x_{a_2b_2}x_{b_2n}|n\rangle}{\langle
1n-1\rangle \langle n~n-2\rangle\langle
n-1|x_{a_2b_2}x_{a_2n}|n\rangle }\frac{\langle
n-1|x_{a_1b_1}x_{b_1n}|n\rangle}{\langle
n-1|x_{a_1b_1}x_{a_1n}|n\rangle }\quad \nn
B_{n;a_1b_1;a_2b_2}^{(1,1)}&=&\frac{\langle 1n\rangle  \langle
n-1|x_{n-1a_2}|n-1]}{\langle 1n-1\rangle  \langle n|x_{na_2}|n-1]
}\frac{\langle n-1|x_{a_1b_1}x_{b_1n}|n\rangle}{\langle
n-1|x_{a_1b_1}x_{a_1n}|n\rangle }\quad (b_2=n-1)\nn
 B_{n;a_1b_1;a_2b_2}^{(1,2)}&=&\frac{\langle 1n\rangle  \langle n-1~b_2-2\rangle (x'_{a_2b_2})^2}{\langle 1n-1\rangle \langle n|x_{na_2}x'_{a_2b_2}|b_2-2\rangle}\frac{\langle n-1|x_{a_1b_1}x_{b_1n}|n\rangle}{\langle n-1|x_{a_1b_1}x_{a_1n}|n\rangle } \nn
 B_{n;a_1b_1;n-3n-1}^{(1,{\rm boundary})}&=&B^{({\rm boundary})}_{n;n-3~n-1}\frac{\langle n-1|x_{a_1b_1}x_{b_1n}|n\rangle}{\langle n-1|x_{a_1b_1}x_{a_1n}|n\rangle },
 \eeqa
where the last term $ B_{n;a_1b_1;n-3n-1}^{(1,{\rm boundary})}$
comes from Eq. (\ref{BNHMV-boundary}). Again the superscripts are
used to show the types of the contributions. For instance, in the superscript $(1,1)$
of $B_{n;a_1b_1;a_2b_2}^{(1,1)}$, the first $``1"$ means that it is the
type-I contribution, while the second $``1"$ implies that it is a descendant from the NMHV
case. A generalization to the N$^k$MHV case will be
$B_{n;a_1b_1;\dots;a_kb_k}^{(m)}$, where $m$ is a string composed of
three kinds of labels, ``$1$" ``$2$" and ``boundary".

As we have mentioned in the NMHV case, and we want to stress it
here again that the bonus coefficients of Fig.~\ref{NNMHV-1} are simply
given as the previous ones, namely the coefficients of NMHV amplitudes,
with replacements $(a_1 \rar a_2, b_1 \rar b_2)$ and
multiplied by the same factor $\frac{\langle
n-1|x_{a_1b_1}x_{b_1n}|n\rangle}{\langle
n-1|x_{a_1b_1}x_{a_1n}|n\rangle }$.

Next, we calculate the contributions from the diagrams in Fig.~\ref{NNMHV-4} which are of the form NMHV$\times$ MHV and we get
\vspace{-.04in}
 \beqa
 \begin{aligned} \label{NNMHV-second}
M_{{\rm II}}=&\sum_{2\leq a_1, b_1\leq n-1}\sum_{a_1\leq a_2, b_2<b_1} \big( B_{n;a_1b_1;a_2b_2}^{(2,1)}\{n;a_1b_1;a_2b_2\}_1(n-1\leftrightarrow b_1-1) \\
&+B_{n;a_1b_1;a_2b_2}^{(2,2)}[\{n;a_1b_1;a_2b_2\}_1(b_2-1\leftrightarrow b_1-1)] \big) \\
&+  \sum_{2\leq a_1\leq
n-3} B_{n;a_1n-1;n-4n-2}^{(2,{\rm boundary})}[\{n;a_1n-1;n-3n-1\}_1(n-2\leftrightarrow
n-1)].
\end{aligned}
\eeqa
In the above sum we do not include the boundary case $(a_1,b_1,a_2,b_2)=(n-4,n-1,n-4,n-2)$, which we shall study separately. The coefficients are given by \beqa
\begin{aligned}
B_{n;a_1b_1;a_2b_2}^{(2,1)}&=\frac{\langle 1n\rangle \langle
n-1~ b_1-2\rangle \langle
n-1|x_{b_2a_2}x'_{b_2b_1}x'_{a_1b_1} x_{a_1n}|n\rangle(x'_{a_1b_1})^2}{\langle
1n-1\rangle\langle b_1-2|x'_{a_1b_1}x_{a_1n}|n\rangle
\langle
n-1|x_{b_2a_2}x'_{a_2b_1}x'_{a_1b_1}x_{a_1n}|n\rangle
}\\
B_{n;a_1b_1;a_2b_2}^{(2,1)}&=\frac{\langle 1n\rangle
\langle n-1|x_{n-1a_2}|n-1](x'_{a_1b_1})^2}{\langle
1n-1\rangle \la n|x_{na_1}x'_{a_1b_1}x'_{b_1a_2}|n-1]
}\quad (b_2=n-2)\\
B_{n;a_1b_1;a_2b_2}^{(2,2)}&=\frac{\langle 1n\rangle   \langle
n-1~b_2-2\rangle(x'_{a_2b_2})^2 (x'_{a_1b_1})^2
}{\langle 1n-1\rangle\langle
n|x_{na_1}x'_{a_1b_1}x'_{b_1a_2}x'_{a_2b_2}|b_2-2\rangle
} \\
B_{n;a_1b_1;n-4n-2}^{(2,{\rm boundary})}&=\frac {\langle 1n \rangle \langle b_1-4~n-1  \rangle [b_1-3~n-1][b_1-2~n-1] (x'_{b_1-3b_1})^2 (x'_{a_1b_1})^2}{ \langle 1~n-1 \rangle [b_1-3~b_1-2]\langle b_1-4|x_{b_1-4~b_1-1}|n-1] \langle n|x_{na_1}x'_{a_1b_1}x_{b_1-1b_1-3}|n-1]}.
\label{coeff4}
\end{aligned}
\eeqa
By comparing the results with those of NMHV, now we are
ready to see the patterns. For this type of diagrams
Fig.~\ref{NNMHV-4}, the bonus coefficients can be obtained from the
results of NMHV by doing the following replacements on the indices
of region momenta $x$'s: $n\rar b_1, a_1 \rar a_2, b_1 \rar b_2$, and $x \rar x'$ when $x$ has the index $n$ with it.
Furthermore one should apply the changes on $\la n|$ and $\la n-i|$, which correspondingly read $\la n |  \rar \la n | x_{na_1}x'_{a_1b_1}$, and $\la n-i| ({\rm or} ~[n-i| ) \rar \la b_1-i| ({\rm or}~ [b_1-i|)$
for $i>1$. Finally we multiply the obtained answers by a factor $(x'_{a_1b_1})^2$.

The bonus coefficients of the contributions from other diagrams are
actually the same as those of the NMHV case. For the sake of completeness, let us
write down these contributions: for the contribution from
Fig.~\ref{NNMHV-3}, we have \beqa M_{{\rm III}}=\sum_{2\leq a_1,
b_1\leq n-1}\sum_{b_1\leq a_2, b_2<n}B_{n;a_1b_1;a_2b_2}^{(2)}[\{n;a_1b_1;a_2b_2\}_2(b_1-1\leftrightarrow
n-1)], \eeqa where the bonus coefficients $B_{n;a_1b_1;a_2b_2}^{(2)}$ are given by Eq.~(\ref{B2}); for the other
contribution coming from Fig.~\ref{NNMHV-2}, we get \beqa M_{{\rm
IV}}=\sum_{2\leq a_1, b_1\leq n-1}\sum_{a_1\leq a_2,b_2<b_1}B_{n;a_1b_1;a_2b_2}^{(1)}\{n;a_1b_1;a_2b_2\}_1, \eeqa and
similarly the coefficients are given by Eq.~(\ref{B1-1}) and
Eq.~(\ref{B1-2}).

Again as in the case of Eq.~(\ref{NNMHV-second}),
this formula does not include the boundary case,
$\{n;a_1b_1;a_2b_2\}_1=\{n;n-4n-1;n-4n-2\}_1$, which should be
considered separately, as we shall do below.

\begin{figure}
  \begin{center}
    \subfigure[ $6$-point diagram deleted by bonus relations]{\label{NNMHV-6-delete}\includegraphics[scale=0.75]{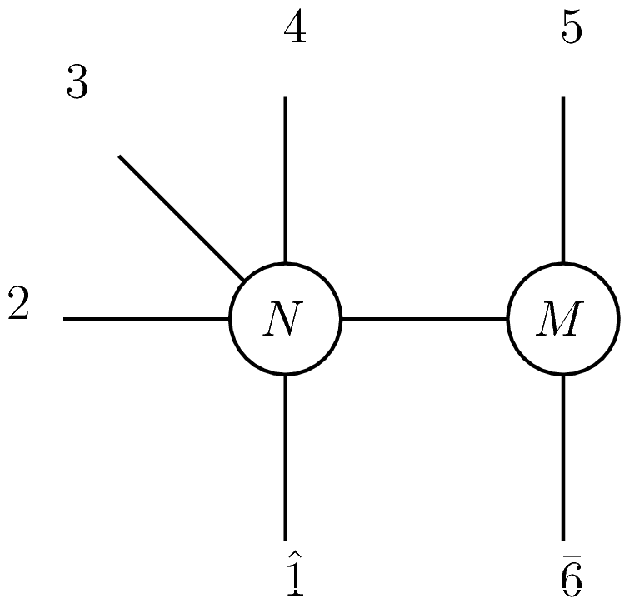}}
    \hspace{1.cm}\subfigure[ $6$-point diagram]{\label{NNMHV-6}\includegraphics[scale=0.75]{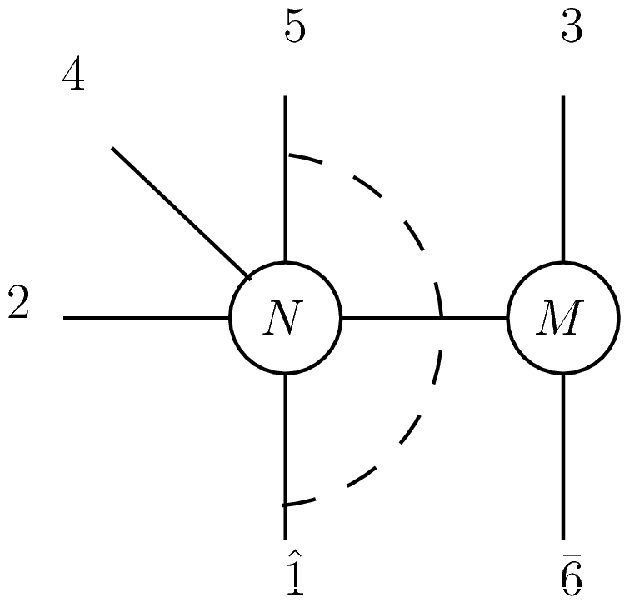}}

  \end{center}
  \caption{ Diagrams for $6$-point N${}^2$MHV amplitude.}
  \label{NNMHV-6-diagrams}
\end{figure}

Similar to 5-point NMHV amplitude, the 6-point N$^2$MHV
amplitude is special which only receives contributions from diagrams of NMHV $\times$ MHV type and we must treat it separately. We can delete Fig.~\ref{NNMHV-6-delete} by bonus relations, and the
contribution from Fig.~\ref{NNMHV-6} gives,
\begin{equation}
\mathcal{M}_6=-\frac{[16][25][45]}{[15][24][56]}
[\{6;25,24\}_1(3\leftrightarrow5)] + \mathcal{P}(2,3,4).
\label{6NN}
\end{equation}
As the NMHV case (\ref{NMHV-boundary}), 6-point N${}^2$MHV
amplitude~(\ref{6NN}) can also be similarly generalized, and we
obtain the boundary term of the full $n$-point N$^2$MHV amplitudes,
\beqa M^{({\rm boundary})}_n &=& B^{(\rm
boundary)}_{n;n-4~n-1;n-4~n-2} [\{n;n-4~n-1;n-4~n-2\}_1
 (n-3\leftrightarrow n-1)],
\eeqa where the bonus coefficients are given as
 \beqa
B^{(\rm boundary)}_{n;n-4~n-1;n-4~n-2} = \frac{\langle 1n\rangle
\langle n-5~ n-1\rangle [n-4~n-1][n-2~n-1] x_{n-4n}^2}{\langle
1n-1\rangle [n-4~n-2]\langle n-5|x_{n-4~n-1}|n-1]\langle
n|x_{n-1~n-4}|n-1]}. \eeqa

Therefore we have calculated all the contributions for N${}^2$MHV
amplitudes and as in the NMHV case, it can also be written as a sum of $(n-3)!$
permutations,
 \beq
 \mathcal{M}_n^{{\rm N}^2{\rm MHV}}=M_{{\rm I}}+M_{{\rm II}}+M_{{\rm III}}+M_{{\rm IV}}+M^{({\rm boundary})}_n+\mathcal{P}(2,3,\ldots ,n-2).
 \label{NNMHVformula}
 \eeq
The result can be proved very similarly by induction as in the NMHV case.

\section{Generalization to all gravity tree amplitudes}

Now we have all the ingredients for generalizing our results and
stating the patterns for all tree-level gravity amplitudes. Our way
of using bonus relations gives the simplified tree-level N$^k$MHV
superamplitude as a sum of $(n-3)!$ permutations, and each of them
contains normal and exchanged contributions, \beq \mathcal{M}^{{\rm
N}^k\rm{MHV}}_n=\big[A_n^{\rm MHV} \big]^2 \big(\sum_{\alpha}
B^{(1,m_1)}_{\alpha} G_{\alpha} R^2_{\alpha} + \sum_{\beta}
B^{(2,m_2)}_{\beta} [G_{\beta} R^2_{\beta}(b_1-1 \leftrightarrow
n-1)] \big)+\mathcal{P}(2,3,\ldots,n-2).\eeq
For both the contributions we have $k$ types of
terms from $k$ BCFW channels, namely N$^p$MHV $\times$ N$^q$MHV, for
$p+q+1=k$ with $0 \leq p, q < k$ by reducing the homogeneous term recursively.
As we have stressed repeatedly, to
respect the ordered structure, we have only used bonus relations on
one lower-point amplitude, namely the right-hand-side N$^q$MHV for
normal contribution, and the left-hand-side N$^p$MHV for exchanged
contribution.

Before presenting all the bonus coefficients for general tree amplitudes,
we pause to show by induction that bonus relations roughly reduce the number
of terms from $(n-2)!$ in the original solution to $(k+1)(n-3)!$ in
the simplified one. To get the previous counting we note that in the N$^p$MHV$\times$ N$^q$MHV channel of the
normal contribution, by applying bonus relations to the N$^q$MHV
lower-point amplitude we can reduce the number of terms from
$(n-2)!/k$ to $(q+1)(n-3)!/k$. Taking into account all channels gives us
$(1+2+\ldots+k)(n-3)!/k$ terms, with the same number from the exchanged
contribution, thus the simplified form has only $(k+1)(n-3)!$ terms.
By parity, one only needs N$^k$MHV amplitudes with $n>2k+2$ legs and
thus the bonus relations can be used to delete at least half of the
terms in tree amplitudes. The simplification becomes more
significant when $n\gg k$.

\begin{figure}[t]
  \begin{center}
    \subfigure[${\rm
MHV} \times {\rm N}^{k-1}{\rm MHV}$  ]{\label{gendiag-1}\includegraphics[scale=0.75]{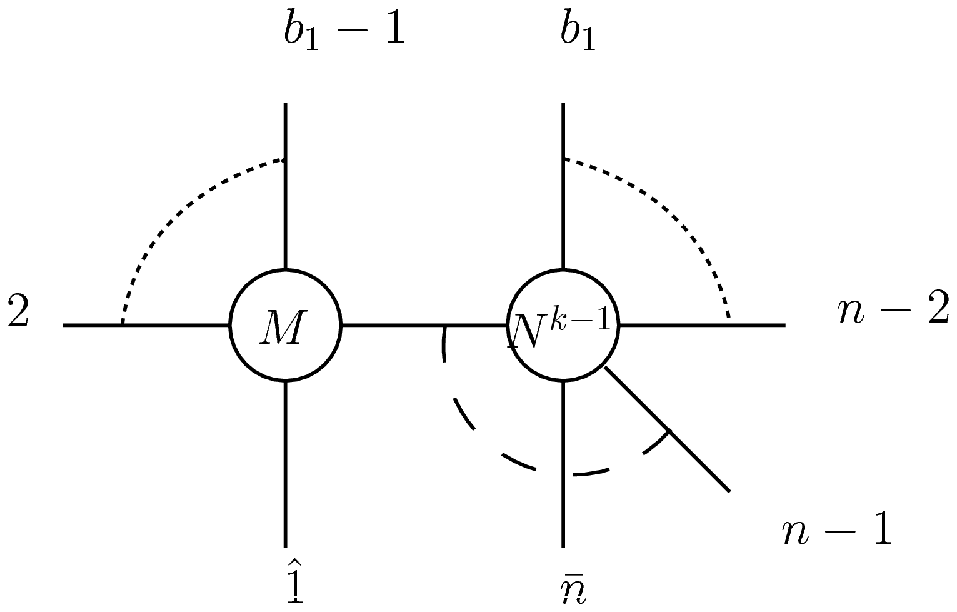}}
    \hspace{1.cm}\subfigure[$ {\rm N}^{k-1}{\rm MHV}\times {\rm
MHV}$ ]{\label{gendiag-2}\includegraphics[scale=0.75]{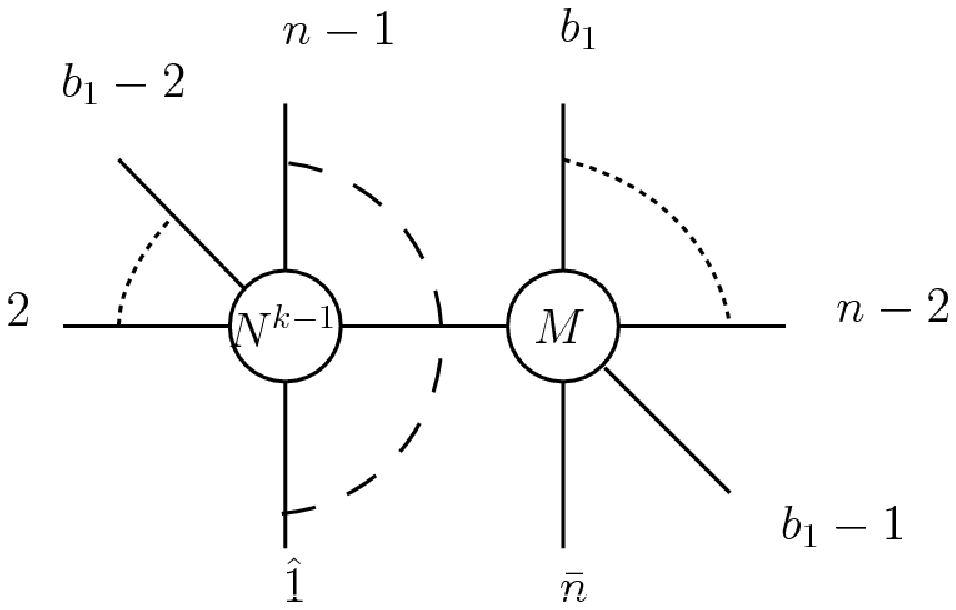}}
\end{center}
  \caption{ Two relevant diagrams for computing new bonus coefficients for $n$-point N${}^{k}$MHV amplitude.
  The rest of the bonus coefficients can be obtained recursively from the N${}^{k-1}$MHV case.}
  \label{gendiagrams}
\end{figure}

Now we generalize the pattern found in the NMHV and N$^2$MHV cases to
write down all the bonus coefficients for general tree amplitudes. As we
have learned from the examples, once the bonus coefficients of
N${}^{k-1}$MHV amplitudes are calculated, then for the N${}^k$MHV
amplitudes, one only needs to compute two types of new contributions
for N$^{k}$MHV amplitudes, namely the normal contribution from ${\rm
MHV} \times {\rm N}^{k-1}{\rm MHV}$ channel ($q=k-1$) and the
exchanged contribution from $ {\rm N}^{k-1}{\rm MHV}\times {\rm
MHV}$ channel ($p=k-1$) (see Fig.~\ref{gendiagrams}). All other bonus
coefficients $B^{(m)}_{\alpha}$ of N$^p$MHV $\times$ N$^q$MHV with
$q<k-1$ and $p<k-1$, are the same as those computed previously,
namely the results from N$^{k-1}$MHV amplitudes. Since the summation
variables of N$^k$MHV amplitude can be obtained by adding a pair of
new labels $a_k,b_k$ to the previous one, $\alpha'$,
$\alpha=\{\alpha';a_k,b_k\}$,
the result can be written as \beqa
B^{(m)}_{\alpha}&=&B^{(m)}_{\alpha'},\eeqa for both normal
contributions with $q<k-1$ and exchanged ones with $p<k-1$.

Thus we only need to calculate two new contributions from
Fig.~\ref{gendiag-1} and Fig.~\ref{gendiag-2}. It is straightforward
to confirm that all the observations we have made for the cases of
NMHV and N$^2$MHV can be directly generalized to all tree-level
amplitudes. First we shall state the rules and then justify them.
Firstly, just like Eq.~(\ref{B1-1}) and Eq.~(\ref{N^2MHV-1}) for
NMHV and N$^2$MHV cases, the bonus coefficients of
Fig.~\ref{gendiag-1}, $B^{(1, m_1)}_{\alpha}$, can be similarly
obtained by the replacements on the indices of the region momenta $x$'s,
$a_i \rar a_{i+1}, b_i\rar b_{i+1}$, for $B^{(m_1)}_{\alpha'}$ of
N$^{k-1}$MHV amplitudes, then multiplying with a simple common
factor of the form $\frac{\langle
n-1|x_{a_1b_1}x_{b_1n}|n\rangle}{\langle
n-1|x_{a_1b_1}x_{a_1n}|n\rangle }$, which are the same for all
tree-level amplitudes, \beq B^{(1, m_1)}_{\alpha}=\frac{\langle
n-1|x_{a_1b_1}x_{b_1n}|n\rangle}{\langle
n-1|x_{a_1b_1}x_{a_1n}|n\rangle }B^{(m_1)}_{\alpha'}(a_i \rar
a_{i+1}, b_i\rar b_{i+1}).\eeq

Secondly, the bonus coefficients for the new exchanged contributions
Fig.~\ref{gendiag-2}, $B^{(2, m_2)}_{\beta}$, can be obtained by
taking $B^{(m_2)}_{\beta'}$ of N$^{k-1}$MHV amplitudes, and performing the
following replacements on the indices of region momenta $x$'s, namely $n
\rar b_1, a_i \rar a_{i+1}, b_i\rar b_{i+1}$, and $x \rar x'$ when
$x$ has index $n$ with it. And for the spinors, we have $\la n |   \rar \la n | x_{na_1}x'_{a_1b_1}$ as
well as $|n-i \ra ({\rm or} ~|n-i] ) \rar |b_1-i \ra ({\rm or}~
|b_1-i])$ for $i>1$. In addition, the obtained answers are further
multiplied by a factor $(x'_{a_1b_1})^2$, \beqa
B^{(2, m_2)}_{\beta}&=&(x'_{a_1b_1})^2 B^{(m_2)}_{\beta'}, \eeqa
where the arguments of $B^{(m_2)}_{\beta'}$ should be changed
under the rules we described above.

All these rules can be understood in a simple way. For the rules of
the normal contributions, the common factor is obtained in the
following way, \beqa(1-\frac{z_i}{z_{n-1}}) \frac{\la n 1 \ra}{\la
n-1 1 \ra} \rar (1-\frac{z_i}{z_{n-1}}) \frac{\la n \widehat{P}
\ra}{\la n-1 \widehat{P} \ra} \rar \frac{\langle
n-1|x_{a_1b_1}x_{b_1n}|n\rangle}{\langle
n-1|x_{a_1b_1}x_{a_1n}|n\rangle },\eeqa where
$(1-\frac{z_i}{z_{n-1}})$ comes from the fact that we delete one
diagram using bonus relations, and $\frac{\la n 1 \ra}{\la n-1 1
\ra}$ is a factor that always appears in every bonus coefficient.

While for the rules of the exchanged contributions, we find that the
factor $(x'_{a_1b_1})^2$ appears because \beqa \la n 1 \ra \rar \la
\widehat{P} \widehat{1} \ra \rar [\widehat{P} \widehat{1}]\la
\widehat{P} \widehat{1} \ra \rar (x'_{a_1b_1})^2, \eeqa and $\la n|
$ changes in the following way under the recursion relations, \beqa
\la n| \rar \la \widehat{P}| \rar \la n 1 \ra [1 \widehat{P}]\la
\widehat{P}| \rar \la n|x_{na_1}x'_{a_1b_1}.\eeqa Besides, the
transformation rule of $x_{n\gamma_i}$ follows as \beqa
x_{n\gamma_i} \rar x_{\widehat{P} \gamma_{i+1}} \rar x'_{b_1
\gamma_{i+1}},\eeqa where $\gamma$ can be $a$ or $b$ and we have
used the fact that $p_{\widehat{P}}=p_{b_1}+\dots + p_{n-2}+
p_{b_1-1}+p_{\widehat{n}}.$ So in this way, we have a complete
understanding of the rules we have proposed.

Finally, as shown in the examples a boundary contribution has to be considered
separately because the special case $(k+4)$-point N$^{k}$MHV
amplitude only has diagrams of N$^{k-1}$MHV $\times$ MHV type. For
this special contribution, it is straightforward to obtain a general
form, \beqa\label{N^kMHV} M^{({\rm boundary})}_n
&=&B^{(\rm{boundary})}_{\beta_0}\big[\big( A_n^{\rm MHV} \big)^2
G_{\beta_0} R^2_{\beta_0} (n-k-1\leftrightarrow n-1)\big], \eeqa
where $\beta_0=\{n;n-k-2~n-1;n-k-2~n-2;\ldots;n-k-2~n-k\}$, and the
coefficients can be written as
 \beqa
\label{N^kMHV} B^{(\rm{boundary})}_{\beta_0}= \frac{\langle 1n\rangle
\langle n-k-3~ n-1\rangle [n-k-2~n-1][n-k~n-1]
x^2_{n-k-2~n}}{\langle 1n-1\rangle [n-k-2~n-2]\langle
n-k-3|x_{n-k-3~n-1}|n-1]\langle n|x_{n-1~n-k-2}|n-1]}. \eeqa
Therefore, we have found a set of explicit rules to write down all the bonus
coefficients for all tree amplitude in $\mathcal{N}=8$ supergravity.

\section{Conclusion and outlook}

In this note, we simplified tree-level amplitudes in $\mathcal{N}=8$
SUGRA, from the BCFW form with a sum of $(n-2)!$ permutations to a
new form as a sum of $(n-3)!$ permutations. This is achieved by
using the bonus relations, which are relations between tree amplitudes
in theories without color ordering. In contrast to the MHV case, a
naive use of the bonus relations ruins the structure of the non-MHV
ordered tree-level solution, thus we proposed an improved
application of the relations, which respects the ordered structure.
The key point here is to apply the bonus relations to only one of two
lower-point amplitudes in any BCFW diagram, which indeed brings
SUGRA amplitudes to a simplified form having a $(n-3)!$-permutation
sum with some bonus coefficients. To illustrate the method, we have explicitly
calculated simplified amplitudes for the NMHV and N$^2$MHV
cases. We have also argued that the pattern generalizes to N$^k$MHV
cases, and presented a simple way for writing down the bonus
coefficients of all amplitudes, thus one can recursively obtain the
simplified form for general SUGRA tree amplitudes.

The simplification is based on an explicit solution from BCFW
recursion relations of SUGRA tree amplitudes
of~\cite{Drummond:2009ge}, which is in spirit similar to but in
details different from KLT relations. From a computational point of
view, any gravity amplitude obtained from $(n-3)!$ (or the newly
proposed $(n-2)!$) form of KLT relations is a sum of $(n-3)!^2$ (or
$(n-2)!^2$) terms; at least in the special case of $\mathcal{N}=8$
SUGRA, an explicit solution with only $(n-2)!$ terms was found
in~\cite{Drummond:2009ge}, which is a significant
simplification\footnote{It would be nice to see if one can derive
the explicit $(n-2)!$ form (similarly our simplified $(n-3)!$ form)
from $(n-2)!$ (similarly $(n-3)!$) KLT relations. For the simplest
MHV case, both have been derived in~\cite{Feng:2010hd}.}.
Furthermore, in this note we have used the bonus relations to reduce
it to a sum with only $(k+1)(n-3)!$ terms. Further simplifications
of gravity tree amplitudes are certainly worth investigating.

Apart from the computational advantages, the simplification is also
conceptually interesting. The relations between gravity and gauge
theories have been reexamined from various perspectives
recently~\cite{Bern:2008qj,BjerrumBohr:2010ta,BjerrumBohr:2010zb} (see also \cite{Nastase:2010xa}). A
common feature, of these ``gravity''$=$``gauge theory''$^2$ methods,
is the freedom of rewriting $(n-2)!$ forms of gravity tree
amplitudes as $(n-3)!$ forms, essentially by using BCJ relations on
the gauge theory side. Our result confirms this freedom at an
explicit level by directly using it to simplify SUGRA amplitudes,
which also suggests that bonus relations may be regarded as explicit
gravity relations induced by Yang-Mills BCJ relations. It may be
fruitful to understand the exact connections between our method,
general forms of KLT relations, and the square relations. In
particular, it would be nice to go beyond SUGRA and see if similar
simplifications occur generally, given that both BCFW recursion
relations and bonus relations are valid in more general gravity
theories.

Bonus relations and simplifications we obtained at tree level can
also have implications for loop amplitudes. Through the generalized
unitarity-cut method, our new form of tree amplitudes can be used in
calculations of loop amplitudes. In addition, the square relations
have been conjectured to hold at loop level~\cite{Bern:2010ue}, thus
we may expect similar simplifications directly for the SUGRA loop
amplitudes.

\begin{acknowledgements}

We are grateful to Y.-t.~Huang, K.~Jin, M.~Spradlin and A.~Volovich for very
helpful conversations. The work of DN and CW was supported in part
by the US Department of Energy under contract DE-FG02-91ER40688 and the US National
Science Foundation under grants PECASE PHY-0643150 and PHY-0548311.

\end{acknowledgements}

\end{document}